\documentclass[sigconf]{acmart}
\usepackage{algorithm}
\usepackage{algpseudocode}
\usepackage{threeparttable}
\usepackage{multirow}
\AtBeginDocument{%
  }
\setcopyright{acmlicensed}
\copyrightyear{2025}
\acmYear{2025}
\acmDOI{XXXXXXX.XXXXXXX}
\acmConference[Preprint]{}{}{}
\begin{document}
\title{KANHedge: Efficient Hedging of High-Dimensional Options Using Kolmogorov-Arnold Network-Based BSDE Solver}

\author{Rushikesh Handal}
\affiliation{%
  \institution{Preferred Networks, Inc.}
  \city{Tokyo}
  \country{Japan}
}

\author{Masanori Hirano}
\email{research@mhirano.jp}
\orcid{0000-0001-5883-8250}
\affiliation{%
  \institution{Preferred Networks, Inc.}
  \city{Tokyo}
  \country{Japan}
}

\renewcommand{\shortauthors}{Handal and Hirano}

\begin{abstract}
  High-dimensional option pricing and hedging present significant challenges in quantitative finance, where traditional PDE-based methods struggle with the curse of dimensionality. The BSDE framework offers a computationally efficient alternative to PDE-based methods, and recently proposed deep BSDE solvers, generally utilizing conventional Multi-Layer Perceptrons (MLPs), build upon this framework to provide a scalable alternative to numerical BSDE solvers. In this research, we show that although such MLP-based deep BSDEs demonstrate promising results in option pricing, there remains room for improvement regarding hedging performance. To address this issue, we introduce KANHedge, a novel BSDE-based hedger that leverages Kolmogorov-Arnold Networks (KANs) within the BSDE framework. Unlike conventional MLP approaches that use fixed activation functions, KANs employ learnable B-spline activation functions that provide enhanced function approximation capabilities for continuous derivatives. We comprehensively evaluate KANHedge on both European and American basket options across multiple dimensions and market conditions. Our experimental results demonstrate that while KANHedge and MLP achieve comparable pricing accuracy, KANHedge provides improved hedging performance. Specifically, KANHedge achieves considerable reductions in hedging cost metrics, demonstrating enhanced risk control capabilities.

\end{abstract}


\maketitle

\section{Introduction}

Financial derivatives, particularly options, are fundamental instruments in modern quantitative finance, providing essential mechanisms for risk management, portfolio optimization, and hedging strategies. These financial contracts grant holders specific rights to trade underlying assets under predetermined conditions, making their accurate valuation and effective hedging critical for financial institutions and market participants \cite{black1973pricing}. The complexity of modern financial markets demands sophisticated approaches to derivative pricing, especially when dealing with high-dimensional options where the computational burden increases substantially with the number of underlying state variables.

The classical approach to option pricing, as established by Black and Scholes \cite{black1973pricing}, formulates the problem as a partial differential equation (PDE). While this framework works well for simple, single-dimensional problems, real-world financial applications often involve high-dimensional scenarios, such as basket options that depend on multiple underlying assets. In such cases, traditional PDE-based methods face the "curse of dimensionality," where computational complexity grows exponentially with the number of dimensions, making direct numerical solutions intractable \cite{bellman1957dynamic}.

To address the curse of dimensionality, the Backward Stochastic Differential Equation (BSDE) formulation has emerged as a powerful alternative approach \cite{pardoux2005backward}. The BSDE framework, connected to PDEs through the Feynman-Kac theorem, transforms the pricing problem into a stochastic optimization problem that avoids explicit computation of high-dimensional Hessian matrices. Traditional numerical methods for BSDEs, such as finite difference schemes (e.g., \cite{teng2021high}) and regression-based approaches (e.g., \cite{Gobet_2005}), have shown some success but still face limitations with respect to convergence when the number of dimensions increases \cite{bellman1957dynamic}, \cite{lemor2006rate}.

Recent advances in machine learning have revolutionized the BSDE approach to high-dimensional option pricing. One such machine learning approach is the Multi-Layer Perceptron (MLP) based methods, part of a family of models called Deep BSDEs, which have shown promising results in pricing high-dimensional options \cite{han2018solving,hure2020deep}, particularly in their ability to directly estimate the option's delta (price sensitivity) either as a direct model output \cite{han2018solving} or by using automatic differentiation capabilities of MLPs \cite{raissi2024forward}, which is essential for dynamic hedging strategies. These deep learning methods can effectively handle the curse of dimensionality by parameterizing the solution functions with neural networks, enabling practical solutions to previously intractable problems.

Other types of architectures like Recurrent Neural Networks (RNNs) \cite{chan2019machine, Kapllani_2024} or Long Short-Term Memory (LSTM) networks \cite{Kapllani_2024} have been utilized to replace standard MLPs to solve the general form of BSDEs. \cite{Kapllani_2024} find that neither RNNs nor LSTMs improve the accuracy of the BSDE solution, i.e., the initial value of the solution corresponding to the underlying PDE (option price in the case of option-related PDEs). Furthermore, they find that in the case of LSTMs, the approximation error increases, which they attribute to the fact that LSTMs violate the Markovian property of BSDEs. Similar arguments can be made about any attention-based model based on \cite{vaswani2023attentionneed} or convolutional neural network based model used in \cite{widianto2023european} that utilize sequences of past information. Hence, in this research we focus on the simpler and often used MLPs as the baseline architecture in Deep BSDEs.

Despite being successful at estimating option prices, Deep BSDEs face significant limitations when it comes to estimating the deltas required for effective hedging. Obtaining precise delta information from neural networks remains problematic regardless of the estimation approach used. Methods that rely on automatic differentiation for delta calculation \cite{raissi2024forward} do not guarantee smooth delta estimates, which is also highlighted in their work. Barring special cases like digital options or barrier options, a smooth delta is required for stable hedging; otherwise leading to gamma spikes. While approaches that directly model deltas using MLPs \cite{han2018solving} offer better control compared to automatic differentiation, the accuracy from the option pricing task does not translate into accurate delta estimation \cite{hientzsch2019introductionsolvingquantfinance}. Thus, efficient hedging is still an open issue when it comes to Deep BSDE solvers.

Compared to MLPs, Kolmogorov-Arnold Networks (KANs), recently introduced in the literature \cite{liu2024kan}, offer a fundamentally different approach to functional approximation based on the Kolmogorov-Arnold representation theorem. Unlike MLPs that rely on fixed activation functions with learnable weights, KANs employ learnable activation functions (typically B-splines) on the edges of the network, providing enhanced flexibility for continuous function modeling. This architectural design is particularly well-suited for delta estimation, as KANs naturally produce smoother derivatives and, more generally, more accurate approximations for continuous functions \cite{liu2024kan}. 

In this work, we introduce \textbf{KANHedge}, a novel method that leverages KANs for solving high-dimensional option pricing and hedging problems via the BSDE methodology. The major architectural difference with MLP-based Deep BSDE solvers such as in \cite{han2018solving} is that at each intermediate time step, we model the option's delta using a KAN. The central research question we aim to answer in this work is: \textbf{Does KANHedge provide improved hedging performance compared to conventional MLP-based BSDE solvers in high-dimensional settings?} 

To answer the research question, we conduct option pricing and hedging experiments using both high-dimensional European and American-style basket options with varying market conditions and dimensions. We find that although both MLP-based Deep BSDE solvers and KANHedge lead to similar option pricing accuracy with pricing errors below $1\%$, the hedging performance of KANHedge is better across all experimental settings. Utilizing the deltas estimated at each time step, we rely on a pure delta hedging strategy to derive the hedging cost distribution. In our set of experiments, KANHedge results in up to $9\%$ improvements across the hedging cost metrics, measured using the conditional value at risk utility function. This improved performance of KANHedge can be attributed to improved delta estimation at each discrete time step.

\section{Related Work}\label{sec:literature}
Option pricing  and hedging have been addressed through several distinct methodological approaches, each with specific advantages and limitations that inform our work. In this section, we focus on related research that combine machine learning aspects with conventional pricing and hedging tasks. 

\textbf{PDE-Based Approaches:} Recent advances in solving high-dimensional PDEs have enabled efficient option pricing beyond traditional finite difference or tree-based methods. The Deep Parametric PDE Method \cite{glau2022deep} enables efficient option pricing across high-dimensional parameter spaces by training neural networks on families of PDEs. Similarly, the Deep Galerkin Method \cite{sirignano2018dgm} provides a mesh-free approach to solve high-dimensional option pricing PDEs using deep learning. However, generally, PDE-based solvers face challenges in ensuring accuracy, generalization, and training efficiency, with empirical studies concluding that BSDE-based methods offer superior price estimates compared to PDE-solvers across various settings \cite{assabumrungrat2023error}. Our proposed KANHedge model falls into this BSDE category, offering the computational advantages of this approach.

\textbf{Direct Hedging Approaches:} Alternative methodologies such as \cite{buehler2019deep, hirano2023adversarial} directly model the hedging output at each time step, enabling estimation of optimal hedging strategies and option prices based on a variety of issuer-specific target utility functions. These approaches can be extended to higher-dimensional settings for basket option pricing and delta computation. However, these studies primarily target European-style options and their applicability to American-style options, where early exercise estimation is crucial in addition to delta computation, is not straightforward. In contrast to these methods, KANHedge, which incorporates criteria for early execution of the option, offers more straightforward applicability to American-style options.

\textbf{KAN-based PDE solvers:} The seminal work in \cite{liu2024kan} demonstrated the potential of KANs in solving PDEs using a simple Poisson equation. \cite{Wang_2025} combined KANs with physics-informed neural networks to solve PDEs in computational mechanics, where they find that KANs significantly outperform MLPs in terms of accuracy. Compared to these PDE solvers, KANHedge enjoys the simplicity of BSDEs and, to the best of our knowledge, is among the early applications of KANs in BSDE solving.

\textbf{KAN-based Option Pricing:} Although KAN architecture is proposed recently, initial studies have begun applying KANs to option pricing. \cite{handal2024kanopdataefficientoptionpricing} replaces the basis functions used in conventional Least Square Monte Carlo (LSMC) methodology with a KAN network to price American-style options. While their approach provides improved price estimates, they only apply it to pricing tasks with a single underlying asset. Furthermore, its applicability as a hedging tool remains unstudied. Compared to their approach, KANHedge extends naturally to higher-dimensional settings and provides both pricing and hedging capabilities through the BSDE framework. 

\section{Foundational Mathematics}\label{sec:math}

This section establishes the mathematical framework underlying our approach, progressing from general PDE formulations to the BSDE methodology that enables our KAN-based solution. 

Following notation from \cite{han2018solving}, we consider a general class of semilinear parabolic PDEs of the form:

\begin{equation}
\frac{\partial u}{\partial t} + \frac{1}{2}\text{tr}(\Gamma\Gamma^T\nabla^2 u) + \mu^T\nabla u + f(x,t,u,\Gamma^T\nabla u) = 0,
\end{equation}

with $t$ denoting time and $T$ as the final time, terminal condition $u(x,T) = g(x)$. Here $u(x,t)$ is the solution, $x \in \mathbb{R}^d$ represents the state variables, $\Gamma(x,t) \in \mathbb{R}^{d \times d}$ is the diffusion matrix, $\mu(x,t) \in \mathbb{R}^d$ is the drift vector and $f$ is the nonlinear term. Note that $x$ varies with time, but for ease of notation we use $x$ instead of $x_t$.

This general formulation encompasses a wide range of European-style financial derivative pricing problems, including those with stochastic volatility, and complex payoff structures. High-dimensional option pricing problems represent a special case of the semilinear parabolic PDEs discussed above. For basket options involving $d$ underlying assets $S_1, S_2, \ldots, S_d$, the option value $V(S_1, S_2, \ldots, S_d, t)$ satisfies the multidimensional Black-Scholes (BS) PDE:
\begin{equation}
\frac{\partial V}{\partial t} + \frac{1}{2}\sum_{i,j=1}^d \sigma_{ij} S_i S_j \frac{\partial^2 V}{\partial S_i \partial S_j} + r\sum_{i=1}^d S_i\frac{\partial V}{\partial S_i} - rV = 0.
\end{equation}
This corresponds to the PDE framework with the specific forms for the individual components given as: 
\begin{itemize}
    \item $x = (S_1, \ldots, S_d)^T$ representing the vector of asset prices
    \item $u(x,t) = V(x, t)$ representing the option value as a function of asset price vector and time
    \item  $\mu(x,t) = r \cdot x$ representing the risk-neutral drift vector with $r$ as the risk-free rate
    \item $\Gamma(x,t) = \text{diag}(x) \cdot \Sigma$ where $\Sigma$ is the volatility matrix and $\sigma_{ij}$ are its elements
    \item $f(x,t,u,\sigma^T\nabla u) = -rV$ representing the linear discounting
    \item $g(x) = \text{Payoff}(x)$ representing the terminal payoff function
    
\end{itemize}

The high-dimensional formulation faces the notorious “curse of dimensionality”, where traditional numerical methods require computational resources that grow exponentially with the number of underlying assets. The main challenge in high-dimensional settings is the computation of the Hessian matrix $\nabla^2 u$, which requires $O(d^2)$ evaluations and becomes prohibitively expensive as $d$ increases.  For basket options with even modest dimensions (e.g., $d \geq 5$), direct numerical solution becomes computationally intractable using conventional finite difference or finite element approaches.

To alleviate this issue, the Feynman-Kac theorem establishes a fundamental connection between semilinear parabolic PDEs and BSDEs. Specifically, if $(Y_t, Z_t)$ is the solution to the BSDE:
\begin{equation}
\begin{aligned}
dY_t &= -f(X_t, t, Y_t, Z_t)dt + Z_t^T dW_t, \\
Y_T &= g(X_T),
\end{aligned}
\end{equation}

where $X_t$ follows the stochastic differential equation:
\begin{equation}
dX_t = \mu(X_t, t)dt + \Gamma(X_t, t)dW_t,
\end{equation}

then $u(x,t) = Y_t$ when $X_t = x$, and $\Gamma^T(x,t)\nabla u(x,t) = Z_t$. Here $W_t$ is a $d$-dimensional standard Brownian motion with correlation matrix $\Lambda$.

This connection is crucial because the BSDE formulation avoids explicit computation of the Hessian and hence, the curse of dimensionality is significantly mitigated.

American options introduce the additional complexity of optimal early exercise. The option holder can choose to exercise at any stopping time $\tau \leq T$, leading to the optimal stopping problem:
\begin{equation}
Y_t = \sup_{\tau \in [t,T]} \mathbb{E}[e^{-r(\tau-t)}h(X_\tau)|\mathcal{F}_t],
\end{equation}

where $h(x)$ is the intrinsic value function. Following the seminal approach by \cite{el1997reflected}, this optimal stopping problem can be formulated as a reflected BSDE:
\begin{equation}
\begin{aligned}
dY_t &= -f(X_t, t, Y_t, Z_t)dt + Z_t^T dW_t + dK_t, \\
Y_T &= g(X_T), \\
Y_t &\geq \Phi(X_t) \text{ for all } t \in [0,T], \\
\int_0^T &(Y_t - \Phi(X_t))dK_t = 0,
\end{aligned}
\end{equation}

where $K_t$ is a non-decreasing process that ensures the constraint $Y_t \geq \Phi(X_t)$. $\Phi(X_t)$ is a Markovian continuous process growth bounded by a polynomial function, and generally $h(X_t)$ can be used as a substitute.  
 In practice, this reflected BSDE can be approximated using a penalty method:
\begin{equation}
dY_t = -f(X_t, t, Y_t, Z_t)dt+ Z_t^T dW_t + \lambda(h(X_t) - Y_t)^+dt,
\end{equation}

where $\lambda > 0$ is a penalty parameter and $(\cdot)^+ = \max(\cdot,0)$.

European options correspond to the case where there is no early exercise constraint. This can be viewed as setting the penalty parameter $\lambda = 0$ in the American option formulation, resulting in the standard BSDE:
\begin{equation}
\begin{aligned}
dY_t &= -f(X_t, t, Y_t, Z_t)dt + Z_t^T dW_t, \\
Y_T &= g(X_T).
\end{aligned}
\end{equation}

This unified framework allows us to handle both European and American options within the same mathematical setting, simply by adjusting the penalty term.

\section{Methodology}\label{sec:method}

This section presents our novel KANHedge method, which integrates KANs into the deep BSDE methodology for high-dimensional option pricing and hedging. We begin by discretizing the BSDE formulation, then detail the neural network approximation strategy, and finally introduce our KAN-based approach that addresses the limitations of traditional MLP methods.

Following the approach established in deep BSDE literature \cite{han2017deep,han2018solving}, we discretize the continuous-time BSDE using an Euler-Maruyama scheme. Consider a uniform time grid:
\begin{equation}
0 = t_0 < t_1 < \cdots < t_N = T, \quad \text{with } \Delta t = \frac{T}{N}.
\end{equation}

The forward stochastic differential equation for the underlying asset prices is discretized as:
\begin{equation}
S_{t_{n+1}} = S_{t_n} + \mu(S_{t_n}, t_n)\Delta t + \sigma(S_{t_n}, t_n)\Delta W_n,
\end{equation}

where $\Delta W_n = W_{t_{n+1}}-W_{t_n} \sim \mathcal{N}(0, \Delta t \cdot I_d)$ are independent Gaussian increments. The general form of BSDE is correspondingly discretized as:
\begin{equation}
\begin{aligned}
Y_{t_{n+1}} &= Y_{t_n} - [f(X_{t_n}, t_n, Y_{t_n}, Z_{t_n}) + \lambda(h(X_{t_n}) - Y_{t_n})^+]\Delta t \\
&\quad + Z_{t_n}^T \Delta W_n.
\end{aligned}
\end{equation}

The key insight of the deep BSDE method proposed in \cite{han2017deep,han2018solving} is to parameterize the unknown quantities using neural networks. Specifically, the initial option value $Y_0 = u_0$ is treated as a trainable scalar parameter and the gradient process $Z_t$ at each time step is approximated by neural networks:
\begin{equation}
Z_{t_n} \approx Z_{\theta_n}(X_{t_n}), \quad n = 0, 1, \ldots, N-1,
\end{equation}

where $Z_{\theta_n}: \mathbb{R}^d \rightarrow \mathbb{R}^d$ are neural networks with trainable parameters $\theta_n$. The training objective minimizes the terminal condition error:
\begin{equation}
\mathcal{L}(\Theta) = \mathbb{E}[(Y_T^{\Theta} - g(X_T))^2],
\end{equation}

where $Y_T^{\Theta}$ is the terminal value obtained by forward simulation using the parameterized networks. This expectation is approximated using Monte Carlo sampling:
\begin{equation}
\mathcal{L_M}(\Theta) \approx \frac{1}{M} \sum_{i=1}^M (Y_T^{\Theta,(i)} - g(X_T^{(i)}))^2,
\end{equation}

where $M$ is the batch size and $(i)$ denotes the $i$-th Monte Carlo path.

\subsection{Traditional MLP-Based Approach}

In existing deep BSDE methods, the delta networks $Z_{\theta_n}$ are typically implemented as MLPs: 
\begin{equation}
Z_{t_n} \approx \text{MLP}_{\theta_n}(X_{t_n}).
\end{equation}
A major architectural point in MLPs is piecewise linear activations like ReLU. KANHedge, described in the following section, leverages the KAN model that provides an alternative to the linear activation functions.  

\subsection{KANHedge}

Our KANHedge method replaces the MLP networks with KANs, which, in theory, offer better functional approximation capabilities for continuous functions. A KAN layer transforms an input $\mathbf{x} \in \mathbb{R}^{n_{\text{in}}}$ to an output $\mathbf{y} \in \mathbb{R}^{n_{\text{out}}}$ via:
\begin{equation}
y_j = \sum_{i=1}^{n_{\text{in}}} \phi_{j,i}(x_i),
\end{equation}

where each $\phi_{j,i}$ is a learnable univariate function, typically implemented as a B-spline:
\begin{equation}
\phi_{j,i}(x) = w_b \cdot b(x) + w_s \cdot \text{spline}(x).
\end{equation}

Here, $b(x)$ is a simple base function (e.g., SiLU), $\text{spline}(x)$ is a B-spline function with learnable control points, and $w_b, w_s$ are learnable weights.

We model the delta networks $Z_{\theta_n}$ with a KAN architecture:
\begin{equation}
Z_{t_n} \approx \text{KAN}_{\theta_n}(X_{t_n}).
\end{equation}

B-spline basis functions in KANs naturally produce smooth outputs, crucial for accurate delta computation. The advantages of KANs for modeling continuous smooth functions have been demonstrated in \cite{liu2024kan}, but they use simpler toy examples for illustration. 

The overall training algorithm for KANHedge is given below:

\begin{algorithm}[htbp]
\caption{KANHedge Training Algorithm}
\label{alg:KANHedge}
\begin{algorithmic}[1]
\State \textbf{Input:} Time steps $N$, batch size $M$, learning rate $\alpha$
\State \textbf{Initialize:} $u_0$, KAN parameters $\{\theta_n\}_{n=0}^{N-1}$
\For{each training epoch}
    \For{each mini-batch}
        \State Sample initial conditions $\{X_0^{(i)}\}_{i=1}^M$
        \State Set $Y_0^{(i)} = u_0$ for all $i$
        \For{$n = 0, 1, \ldots, N-1$}
            \State Compute $Z_{t_n}^{(i)} = \text{KAN}_{\theta_n}(X_{t_n}^{(i)})$
            \State Generate $\Delta W_n^{(i)} \sim \mathcal{N}(0, \Delta t \cdot \Lambda)$
            \State Update $X_{t_{n+1}}^{(i)} = X_{t_n}^{(i)} + \mu \Delta t + \sigma \Delta W_n^{(i)}$
            \State Update $Y_{t_{n+1}}^{(i)} = Y_{t_n}^{(i)} - f \Delta t + (Z_{t_n}^{(i)})^T \Delta W_n^{(i)}$ \\
            \hspace*{3.5cm} $+ \lambda(h(X_{t_n}^{(i)}) - Y_{t_n}^{(i)})^+ \Delta t$
        \EndFor
        \State Compute MSE loss: $\mathcal{L} = \frac{1}{M} \sum_{i=1}^M (Y_T^{(i)} - g(X_T^{(i)}))^2$
        \State Update parameters: $\Theta \leftarrow \Theta - \alpha \nabla_\Theta \mathcal{L}$
    \EndFor
\EndFor
\State \textbf{Return:} Trained parameters $\Theta^*$ and $u_0$
\end{algorithmic}
\end{algorithm}

\section{Experimental Setup}\label{sec:experiments}

We evaluate the proposed KANHedge method through European and American-style basket option experiments designed to test both pricing accuracy and hedging performance in high-dimensional basket option settings. 
The choice of basket options allows us to systematically increase dimensionality while maintaining economic relevance, as basket options are widely traded instruments in equity and commodity markets.

\subsection{European Basket Options}\label{subsec:experiment_eu}

For our first experimental setup, we consider equal-weight geometric European basket call options following the framework established by \cite{chen2021deep}. The geometric basket construction provides several analytical advantages that make it particularly suitable for validating our KANHedge methodology.

\textbf{Problem Formulation:}
We consider a basket of $d$ assets with price processes $S_1, S_2, \ldots, S_d$, with equal volatility $\sigma$, that follow correlated geometric Brownian motions such that $d\langle W^{(i)}, W^{(j)} \rangle_t = \rho_{ij} dt$; with $\rho_{ij} = \rho$ for $i\neq j$ and $1$ otherwise. The risk-free rate is given as $r_f$. The geometric basket is defined as:
\begin{align}
G_t = \prod_{i=1}^d (S_i)^{1/d},
\end{align}
The payoff at maturity $T$ is given by $\Phi(S_T) = \max\left(G_T - K, 0\right)$.

\textbf{Analytical Solution and Benchmarking:}
A key advantage of this basket option is that it admits an analytical solution under the BS framework \cite{black1973pricing}. The geometric basket $G_t$ follows a single geometric Brownian motion with equivalent volatility $\sigma_G$ and an equivalent risk-free rate $r_{G}$ \cite{chen2021deep}:
\begin{equation}
\begin{aligned}
\sigma_G^2 &= \frac{1+(d-1)\rho}{d} \sigma^2, \text{and} \\
r_G &= r_f - \frac{1}{2}(\sigma_G^2 - \sigma^2).
\end{aligned}
\end{equation}

This allows us to compute exact option prices using the standard BS formula.

\textbf{Experimental Configuration:}
To systematically evaluate the performance of our KANHedge, we design a controlled experimental setup with a baseline configuration and targeted parameter variations. Our baseline configuration consists of a 10-dimensional geometric basket call option with the following market parameters: initial asset prices $S_i^0 = 10$ for all $i = 1, \ldots, d$, $\sigma = 0.1$, $r_f = 0.01$,  $\rho = 0.3$ and an at-the-money strike price $K = 10$.

To assess the robustness and sensitivity of our approach, we systematically vary individual market parameters. Specifically, we examine:
\begin{itemize}
\item Increased market volatility with $\sigma_i = 0.2$ to test performance under higher uncertainty
\item Modified correlation structure to evaluate sensitivity to asset dependencies
\item Out-of-the-money options with $K = 11$ to assess performance across different moneyness levels
\item Higher dimensionality scenarios to validate scalability properties
\end{itemize}

This experimental design allows us to isolate the impact of each market parameter on our method's pricing and hedging performance. 

\subsection{American Basket Options}\label{subsec:experiment_am}

Our second experimental study focuses on American-style basket options, following the setup of \cite{hanbali_am_op}. American options present additional computational challenges due to the optimal stopping problem, making them ideal for testing the robustness of our KANHedge method in complex scenarios without analytical solutions.

\textbf{Problem Formulation:}
We consider American put options on arithmetic average baskets with payoff:
\begin{align}
\Phi(S_t) = \max\left(K - \frac{1}{d}\sum_{i=1}^d S_i, 0\right).
\end{align}

Following \cite{longstaff2001valuing}, we employ LSMC with polynomial basis functions up to order 3 to estimate reference option prices. To maintain computational tractability in high-dimensional settings, our LSMC implementation uses the arithmetic mean of all stock prices $\bar{S}_t = \frac{1}{d}\sum_{i=1}^d S_i$ as the primary modeling variable, rather than considering all combinations of individual asset prices. This dimensionality reduction approach significantly reduces the complexity of the LSMC regression for high-dimensional basket options while maintaining sufficient accuracy for benchmarking purposes.

\textbf{Experimental Configuration:}
The American basket experiments follow the benchmark configuration from \cite{hanbali_am_op}, using an equally weighted basket of $d = 8$ assets. The experimental parameters are:
\begin{itemize}
\item Initial asset prices: $S_i^0 = 4.0$ for all $i = 1, \ldots, 8$
\item Volatilities: $\sigma_1 \in \{0.3, 0.9\}$ for the first asset, $\sigma_2 = 0.6$, $\sigma_3 = 0.1$, $\sigma_4 = 0.9$, $\sigma_5 = 0.3$, $\sigma_6 = 0.7$, $\sigma_7 = 0.8$, $\sigma_8 = 0.2$
\item Risk-free rate: $r_f = 0.01$, Time to maturity: $T = 1$ year
\item Correlation structure: $\rho_{ij} \in \{0.3, 0.8\}$ for $i \neq j$, $\rho_{ii} = 1$
\item Strike prices: $K \in \{3.5, 4.0, 4.5\}$
\end{itemize}

This comprehensive parameter grid allows us to evaluate our method across different market conditions: varying levels of volatility heterogeneity (through the first asset's volatility), different correlation regimes, multiple moneyness levels, and various time horizons.

The positive risk-free rate ($r_f = 0.01$) is particularly important for American put options as it creates conditions where early exercise becomes optimal when the basket value falls sufficiently below the strike price. This occurs because the discounted value of receiving the strike price immediately can exceed the expected discounted payoff from holding the option until maturity. This early exercise feature provides a rigorous test of our method's ability to capture optimal stopping decisions based on the current basket value, which is essential for accurate American option pricing and hedging.

Our experimental design deliberately uses a constant interest rate model, which is simpler than stochastic interest rate frameworks such as the Cox-Ingersoll-Ross (CIR) model \cite{cox1985theory}. This choice allows us to isolate the performance of our KAN-based approach on the core challenge of high-dimensional American option pricing without the additional complexity of stochastic interest rates.

\subsection{Evaluation criteria}\label{subsec:evaluation}

Our evaluation framework encompasses both pricing accuracy and hedging performance. Pricing accuracy enables it to compare the option value estimated by a model against a benchmark, which is the same metrics used in \cite{chen2021deep, hanbali_am_op}. Hedging performance, on the other hand, allows it to compare deltas estimated by each model across all time steps and paths. All evaluation metrics are computed on out-of-sample data using 10,000 simulated paths for the underlying assets to ensure robust statistical assessment of model performance and generalization capabilities.

\textbf{Pricing Evaluation:}
Pricing accuracy serves as the fundamental metric for validating our KANHedge method.  We assess the accuracy using the pricing error, which is measured as the percentage difference between the model price estimate $u_{0}^{M}$ and the target price $u_{0}^{T}$ as:
\begin{align}
\text{Price Error} = 100\cdot\frac{|u_{0}^{M} - u_{0}^{T}|}{u_{0}^{T}}.
\end{align}
For the European geometric basket, we leverage the availability of analytical BS solutions to compute the exact option price as the target. For American basket options, where no analytical solutions exist, we use LSMC as the reference pricing method. While LSMC introduces its own approximation error, it provides a widely accepted benchmark for American option pricing in the literature \cite{hanbali_am_op, chen2021deep}. For convergence of LSMC we use 100,000 simulated paths.

\textbf{Hedging Cost Analysis:}
Beyond pricing accuracy, the quality of hedging strategies is crucial for practical applications. Following a similar methodology to that of \cite{negyesi2023deep, buehler2019deep, hirano2023adversarial}, we evaluate hedging performance through the hedging cost, which measures the actual cost of maintaining a hedged portfolio. From the option issuer's perspective, the option premium is the value that mitigates the cost of hedging, calculated using some issuer-specific utility function across the hedging cost distribution, and a more efficient hedger leads to a smaller hedging cost \cite{buehler2019deep}. 

Similar to \cite{buehler2019deep, hirano2023adversarial}, the hedging cost focuses solely on the rebalancing costs, cash account evolution, and final payoff settlement. Crucially, this evaluation methodology ensures that model comparison is based exclusively on the quality of delta estimates produced by each model, rather than on the price estimate from the BSDE. This separation provides a direct assessment of a model's capability as a hedging instrument, independent of its option valuation performance. The hedging cost $C$ is computed as:
\begin{equation}
\begin{aligned}
C =  e^{-r\tau}\left[- \sum_{n=1}^{N} \delta^{M}_{t_n} (S_{t_{n}} - S_{t_{n-1}})^{T} - e^{r \Delta t}\cdot D_{t_{n-1}} + \Phi(S_\tau) \right],
\end{aligned}
\end{equation}
where:
\begin{itemize}
\item $\Delta t$ is the discrete time step and $\tau$ is the exercise time $\tau$
\item The first term captures the cost of delta rebalancing across all assets
\item The second term accounts for the interest earned/paid on the cash account
\item The third term represents the option payoff at exercise time $\tau$ (at maturity $T$ for European options, or at optimal early exercise time for American options)
\item The entire hedging cost is discounted from exercise time $\tau$ to initial time $t_0$ using the factor $e^{-r\tau}$, ensuring that costs from different samples and exercise times are comparable on a present value basis
\item After early exercise (i.e., after $\tau$), $\delta$ becomes $0$ and no cash flows occur in the cash account
\end{itemize}

Using the hedging cost distribution across all the evaluation sample paths, various target risk measures such as mean, CVaR (Conditional Value at Risk), Entropic Risk Measure (ERM), etc., can be applied to get the optimal option price \cite{buehler2019deep}. A better hedging strategy should result in a lower hedging cost metrics across these risk measures. 

In this research, to assess the tail risk in hedging performance, we employ CVaR at the $95\%$ confidence level for the hedging cost comparison. This measure implies an option issuer who wants to mitigate their losses in the tail region rather than over the entire distribution. With the Value at Risk (VaR) at confidence level $\alpha$ given as:
\begin{equation}
\text{VaR}_{\alpha} = \inf\{x \in \mathbb{R} : P(C > x) \leq 1-\alpha\},
\end{equation}
the CVaR, normalized by the target option price to ensure scale invariance is given as: 
\begin{equation}
\begin{aligned}
{CVaR}_{\alpha} = \frac{\mathbb{E}[C | C \geq \text{VaR}_{\alpha}]}{|u_{0}^{T}|},
\end{aligned}
\end{equation}
where $u_{0}^{T}$ is the BS option price for European options or the LSMC reference price for American options. This normalization allows for meaningful comparison across different option values and market conditions.

\section{Results}\label{sec:results}

We present comprehensive experimental results evaluating KANHedge model against traditional MLP-based approaches across both European geometric basket options and American basket options.

\subsection{European Geometric Basket Options}

Table~\ref{tab:european_results} presents results for European geometric basket option experiments with setup provided in Section \ref{subsec:experiment_eu}, where we can leverage analytical BS solutions for exact benchmarking. 

\begin{table}[htbp]
\centering
\caption{European Geometric Basket Option Results}
\label{tab:european_results}
\begin{threeparttable}
\begin{tabular}{@{}lcccc@{}}
\toprule
Configuration & Method & Price Error (\%) & CVaR \\
\midrule
\multirow{2}{*}{Baseline} & MLP &0.314  & 1.438  \\
      & KANHedge & \textbf{0.055}  & \textbf{1.409} \\
\midrule
\multirow{2}{*}{High Volatility} & MLP & 1.212  & 1.397  \\
       & KANHedge & \textbf{0.943}  & \textbf{1.350} \\
\midrule
\multirow{2}{*}{High Correlation} & MLP & \textbf{0.849} & 1.376
 \\
       & KANHedge & 0.864  & \textbf{1.318} \\
\midrule
\multirow{2}{*}{Out-of-Money} & MLP & 0.912 & 1.819  \\
       & KANHedge & \textbf{0.685}  & \textbf{1.787} \\
\midrule
\multirow{2}{*}{Higher dimension} & MLP & 0.998  & 1.911  \\
       & KANHedge & \textbf{0.831}  & \textbf{1.883} \\
\bottomrule
\end{tabular}
\begin{tablenotes}
\small
\item Price errors are relative to analytical BS solutions. CVaR evaluates normalized hedging cost at 95\% confidence level. Bold indicates best performance.
\end{tablenotes}
\end{threeparttable}
\end{table}
The pricing errors for both models are in the range of $1\%$ for all sets of variations, indicating that $u_{0}$ is roughly the same irrespective of the model used. The distinction comes in the hedging cost where KANHedge outperforms the MLP model, with improvements ranging from $2.01\%$ for the baseline case and increasing up to $4.25\%$ under increased volatility. 

\begin{figure}[htbp]
\centering
\includegraphics[width=\columnwidth]{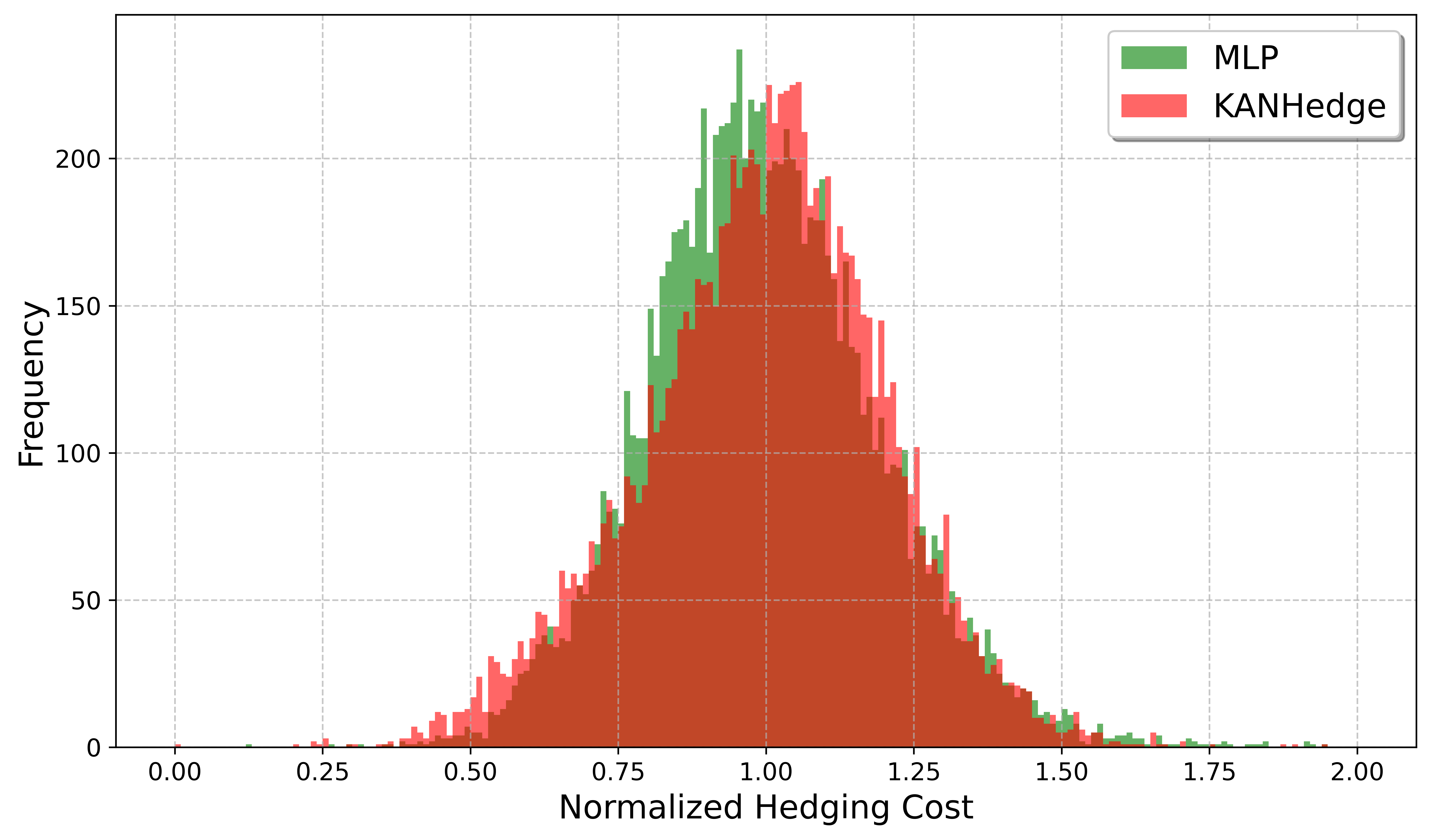}
\Description{Normalized Hedge cost distribution for baseline experiment}
\caption{Normalized Hedging Cost Distribution for European Geometric Basket Option: Baseline. KANHedge (red) shows slightly better hedging cost compared to MLP (green) when tail risk is used as hedging cost.}
\label{fig:hedge_cost_eu}
\end{figure}

We illustrate the hedging cost distribution for the baseline in Figure \ref{fig:hedge_cost_eu}. Although the distributions mostly overlap for the two models, the MLP model has a slightly wider right tail leading to a higher hedging cost CVaR. For the sake of brevity, we only show the distribution for the baseline case, but the overall trend is similar for other sets of experimental variations.

\subsection{American Basket Options}

Table~\ref{tab:american_results} presents results for American basket options following the experimental configuration described in Section~\ref{subsec:experiment_am}. Since no analytical solutions exist, pricing accuracy is measured against LSMC benchmarks.
\begin{table}[htbp]
\centering
\caption{American Basket Option Results}
\label{tab:american_results}
\begin{threeparttable}
\begin{tabular}{@{}lccc@{}}
\toprule
Configuration & Method & Price Error (\%) & CVaR \\
\midrule
\multirow{2}{*}{$\sigma_1=0.3$, $\rho=0.3$, $K=35$} & MLP & 0.586 & 1.664 \\
                                                     & KANHedge & \textbf{0.374} & \textbf{1.522} \\
\midrule
\multirow{2}{*}{$\sigma_1=0.3$, $\rho=0.3$, $K=40$} & MLP & 0.257 & 1.540 \\
                                                     & KANHedge & \textbf{0.108} & \textbf{1.397} \\
\midrule
\multirow{2}{*}{$\sigma_1=0.3$, $\rho=0.3$, $K=45$} & MLP & 0.551 & 1.602 \\
                                                     & KANHedge & \textbf{0.172} & \textbf{1.486} \\
\midrule
\multirow{2}{*}{$\sigma_1=0.3$, $\rho=0.8$, $K=35$} & MLP & 0.279 & 1.479 \\
                                                     & KANHedge & \textbf{0.080} & \textbf{1.376} \\
\midrule
\multirow{2}{*}{$\sigma_1=0.3$, $\rho=0.8$, $K=40$} & MLP & \textbf{0.213} & 1.416 \\
                                                     & KANHedge & 0.352 & \textbf{1.330} \\
\midrule
\multirow{2}{*}{$\sigma_1=0.3$, $\rho=0.8$, $K=45$} & MLP & \textbf{0.004} & 1.374 \\
                                                     & KANHedge & 0.370 & \textbf{1.280} \\
\midrule
\multirow{2}{*}{$\sigma_1=0.9$, $\rho=0.3$, $K=35$} & MLP & 0.174 & 1.441 \\
                                                     & KANHedge & \textbf{0.093} & \textbf{1.334} \\
\midrule
\multirow{2}{*}{$\sigma_1=0.9$, $\rho=0.3$, $K=40$} & MLP & 0.327 & 1.359 \\
                                                     & KANHedge & \textbf{0.301} & \textbf{1.281} \\
\midrule
\multirow{2}{*}{$\sigma_1=0.9$, $\rho=0.3$, $K=45$} & MLP & \textbf{0.231} & 1.338 \\
                                                     & KANHedge & 0.426 & \textbf{1.225} \\
\midrule
\multirow{2}{*}{$\sigma_1=0.9$, $\rho=0.8$, $K=35$} & MLP & \textbf{0.149} & 1.286 \\
                                                     & KANHedge & 0.154 & \textbf{1.216} \\
\midrule
\multirow{2}{*}{$\sigma_1=0.9$, $\rho=0.8$, $K=40$} & MLP & 0.089 & 1.347 \\
                                                     & KANHedge & \textbf{0.035} & \textbf{1.250} \\
\midrule
\multirow{2}{*}{$\sigma_1=0.9$, $\rho=0.8$, $K=45$} & MLP & \textbf{0.083} & 1.282 \\
                                                     & KANHedge & 0.153 & \textbf{1.225} \\
\bottomrule
\end{tabular}
\begin{tablenotes}
\small
\item Price errors are relative to LSMC reference values. CVaR evaluates normalized hedging cost at 95\% confidence level. Bold indicates best performance.
\end{tablenotes}
\end{threeparttable}
\end{table}

A similar trend is observed in the case of American basket options where pricing errors for both models are within $1\%$, with no clear advantage to either model. As for the hedging cost, KANHedge outperforms the MLP model in all experimental variations with improvements ranging from $4.44\%$ to $9.28\%$. The larger improvements in the case of American options can be attributed to the increased volatility levels across the assets in the basket. 

\begin{figure}[htbp]
\centering
\includegraphics[width=\columnwidth]{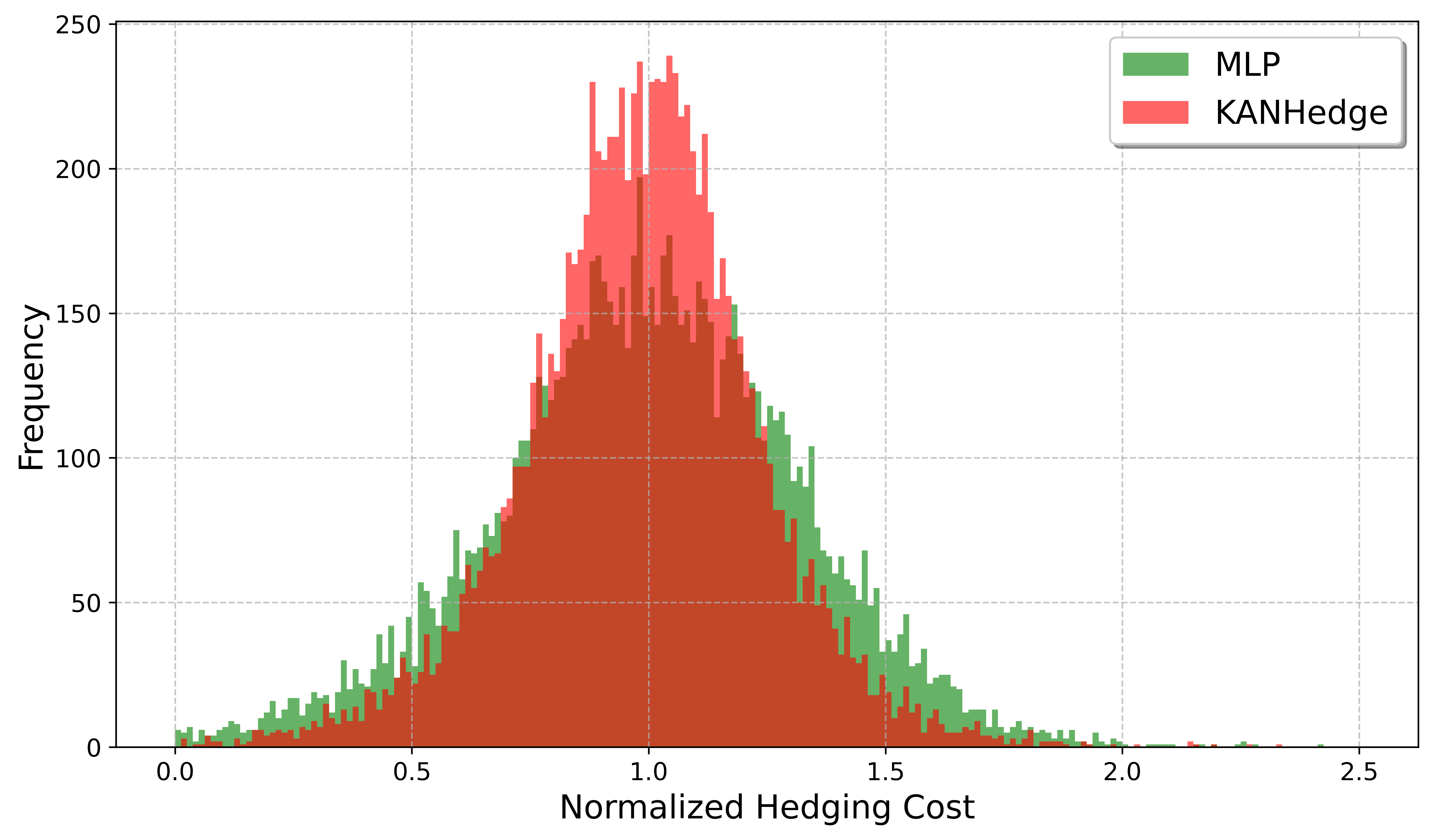}
\Description{Normalized Hedge cost distribution for $\sigma_1=0.3$, $\rho=0.3$, $K=35$}
\caption{Normalized Hedging Cost Distribution for American Basket Option: $\sigma_1=0.3$, $\rho=0.3$, $K=35$. KANHedge (red) shows better hedging cost compared to MLP (green) when tail risk is used as hedging cost.}
\label{fig:hedge_cost_am}
\end{figure}

We illustrate the hedging cost distribution for the $1^{st}$ experiment with  $\sigma_1=0.3$, $\rho=0.3$, $K=35$ in Figure \ref{fig:hedge_cost_am}. KANHedge shows a clear improvement over MLP with a more concentrated hedging cost in the tail region. Here too, for the sake of brevity, we only show the hedging cost distribution for a single variation, but the overall trend is the same across other combinations of parameters.

\section{Discussion}\label{sec:discussion}

The experimental results reveal important insights into the differences between KANHedge and MLP architectures for option pricing and hedging applications, particularly highlighting the distinction between pricing accuracy and hedging quality.

The comparable pricing accuracy between KANHedge and MLP suggests that both architectures possess sufficient representational capacity to learn the underlying option valuation. This finding indicates that the choice between KANHedge and MLP should not be primarily driven by pricing accuracy considerations, as both methods can effectively approximate the complex high-dimensional pricing functions inherent in basket options.

However, the consistently improved hedging cost performance of KANHedge reveals a more nuanced architectural advantage. The CVaR improvements across all experiments demonstrate that KANHedge results in better gradient estimation capabilities. This distinction is critical for practical applications, where different option issuers can have different utility functions to be optimized over the hedging cost distribution. Although in our research we use CVaR as the target utility, a more concentrated distribution, as is the case in Figure \ref{fig:hedge_cost_am}, should result in lower hedging cost metrics over other utility functions such as Value at Risk or ERM.

\begin{figure}[htbp]
\centering
\includegraphics[width=\columnwidth]{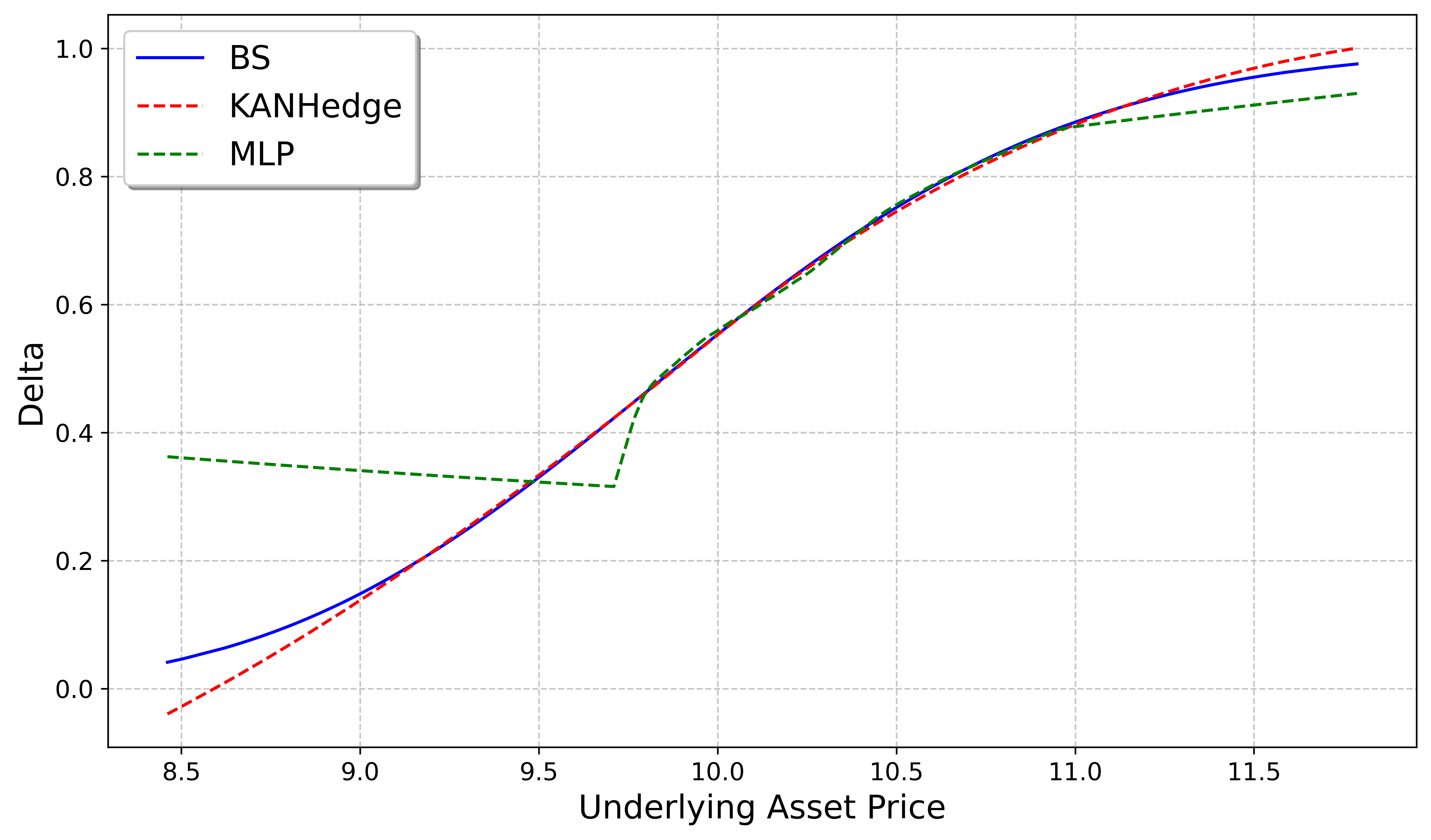}
\Description{Deltas estimated by KANHedge and MLP model compared against BS model delta for single dimension European option}
\caption{Delta estimation for single-dimensional European call option. KANHedge (red) shows superior approximation to the analytical Black-Scholes delta (blue) compared to MLP (green), particularly in out-of-the-money regions.}
\label{fig:kan_vs_mlp_delta}
\end{figure}

The improved delta estimation of KANHedge can potentially be attributed to the smoother and learnable activation functions in the architecture. B-spline basis functions used in KANs have the ability to produce smoother and more accurate derivatives compared to MLPs with piecewise linear activations like ReLU. To illustrate this advantage, we consider a variant of the geometric basket option experiments; we change the number of dimensions to 1 in the baseline case of the experiments. The resulting option is nothing but a simple European option. The resulting delta at some random intermediate time step $t_{10}$ is given in Figure \ref{fig:kan_vs_mlp_delta}.

The advantage of KANHedge in modeling a smooth delta function becomes clear in this simple European option setting. Prices estimated by both models are within $1\%$ of the BS price, but deltas produced by KANHedge align closely with the BS delta in this simple experiment. Although the BS delta is only near-optimal in the discrete time setting, smaller volatility, as is the case in the example, makes the BS delta a good indicator of the optimal delta.

To focus solely on the gradient modeling capabilities of KANHedge, we rely on a delta hedging approach. The smoother and more accurate delta estimation indicates that KANHedge can potentially lead to better gamma estimates for options. We leave pricing and delta-gamma hedging of portfolios of multiple options as a future research direction. 

\section{Conclusion}\label{sec:conclusion}

In this study, we proposed KANHedge, a previously unexplored method that leverages Kolmogorov-Arnold Networks for high-dimensional option pricing and hedging via backward stochastic differential equations. The key idea in this research is to replace traditional MLPs, used to model the option's delta at each discrete time step, with KANs, which employ learnable B-spline activation functions, thus providing enhanced function approximation capabilities for continuous derivatives. Through comprehensive experiments on both European and American basket options, we demonstrated that while KANHedge and MLP achieve comparable pricing accuracy, KANHedge provides improved delta estimation resulting in improved hedging capabilities. Our experimental results reveal that KANHedge achieves lower hedging cost metrics, measured with CVaR, with improvements up to $4\%$ for European basket options and up to $9\%$ for American basket options.

\bibliographystyle{ACM-Reference-Format}
\bibliography{kanhedge}

\end{document}